\documentclass[aps,prd,reprint,nofootinbib,longbibliography,showkeys,dvipsnames, onecolumn]{revtex4-2}
\pdfoutput=1
\usepackage{geometry,amsmath,amsfonts}
\usepackage{slashed}
\usepackage{epsfig}
\usepackage{latexsym}
\usepackage{graphicx}
\usepackage[justification=RaggedRight]{caption}
\usepackage{amssymb}
\usepackage{subfig}
\usepackage{color}
\usepackage{multirow}
\usepackage{color}
\usepackage{rotating}
\usepackage{ifthen}
\usepackage{epsfig}
\usepackage{hyperref}

\newcommand{\eq}{\mathrm{eq}}
\newcommand{\CP}{\mathcal{CP}}

\begin{document}

\title{Baryogenesis and Dark Matter from non-thermally produced WIMPs}
\author{Giorgio Arcadi$^{a,b}$}
\email{giorgio.arcadi@unime.it}

\author{Sarif Khan $^{c}$}
\email{skhan.phys@aliah.ac.in}

\author{Agnese Mariotti $^{d}$}
\email{agnese.mariotti@itp.uni-hannover.de}

\vspace{0.1cm}
 \affiliation{
${}^a$ 
 Dipartimento di Scienze Matematiche e Informatiche, Scienze Fisiche e Scienze della Terra, Universita degli Studi di Messina, Via Ferdinando Stagno d'Alcontres 31, I-98166 Messina, Italy
}

\vspace{0.1cm}
 \affiliation{
${}^b$ 
INFN Sezione di Catania, Via Santa Sofia 64, I-95123 Catania, Italy
}

\vspace{0.1cm}
\affiliation{${}^c$ Department of Physics, Aliah University,
	Kolkata 700160, India}

\vspace{0.1cm}
 \affiliation{${}^d$ 
Institut für Theoretische Physik, Leibniz Universität Hannover, Appelstraße 2, 30167 Hannover, Germany
}

\begin{abstract} 
We illustrate, via a simplified model, a scenario in which the baryon-asymmetry and, possibly the dark matter component of the Universe are simultaneously generated by the decay of a WIMP-like mother particle, in turn produced non-thermally during an epoch of Early Matter domination.
We first consider the standard evolution of the Universe and introduce TeV-scale BSM particles, finding that this paradigm cannot produce enough baryon asymmetry.
% We first investigate the standard evolution of the Universe, finding that we cannot obtain the correct value of the baryon asymmetry by introducing TeV-scale BSM particles.
This deficiency can be resolved by considering a non-standard scenario, with a matter-dominated phase prior to radiation-domination. Finally, we include a dark matter candidate, which is non-thermally produced during the Early Matter domination. Our results demonstrate an interesting common origin of baryon asymmetry and Dark Matter, with the particle masses lying within the collider-detectable range, thanks to the presence of non-standard evolution in the early Universe.
\end{abstract}

\maketitle

\section{Introduction}

Understanding why the visible Universe is made only of ordinary baryonic matter, with a substantial absence of antimatter of primordial origin, is one of the most stimulating challenges of modern astroparticle physics. The most commonly accepted explanation for this observational evidence is the existence of a dynamical mechanism, generally dubbed baryogenesis, which generated an asymmetry between baryon and anti-baryons, starting from symmetric initial conditions in the primordial Universe. Over the years, many baryogenesis mechanisms have been proposed, all complying with a set of necessary conditions established by Sakharov~\cite{Sakharov:1967dj}. Among them, baryogenesis via leptogenesis~\cite{Fukugita:1986hr,Covi:1996wh,Buchmuller:2004nz,Davidson:2008bu} and Electroweak baryogenesis~\cite{Kuzmin:1985mm,Shaposhnikov:1986jp,Farrar:1993hn} are very popular as they establish an interesting connection between baryogenesis and, respectively, the origin of neutrino masses and Higgs physics.

An interesting alternative is represented by a framework in which baryogenesis follows a similar dynamics as the WIMP-mechanism~\cite{Cui:2015eba}. In this scenario, the baryon asymmetry is generated via out-of-equilibrium $B$- and $\CP$-violating annihilations and/or decays of a thermal relic with mass in the $\mathcal{O}(100\,\mbox{GeV})-\mathcal{O}(100\,\mbox{TeV})$ range. The key advantage of this framework is that the baryon abundance can be related to experimentally testable particle physics inputs, in analogy with the WIMP paradigm. For proposals along this reasoning, the interested reader might look, for example, at~\cite{McDonald:2011sv,Baldes:2014gca,Baldes:2014rda,Baldes:2015lka,Cui:2012jh,Cheung:2013hza,Cui:2013bta,Rompineve:2013grm,Davoudiasl:2015jja,Cui:2011ab,Bernal:2012gv,Arcadi:2013jza}.

A further advantage is the fact that dark matter (DM) production can be easily incorporated in such a setup. Refs.~\cite{Arcadi:2013jza} and \cite{Arcadi:2015ffa} presented a scenario in which DM and baryon abundances are simultaneously produced via out-of-equilibrium decay of a thermally-produced WIMP-like particle. However, this setup has the unappealing feature of requiring a rather heavy mass spectrum, hardly accessible by Earth-based experiments, in order to reproduce the correct values of both the DM relic density and the baryon asymmetry. This outcome stems from the necessity of a large thermal abundance for the mother particle in order to compensate for the typically tiny $\CP$-asymmetry generated in its decay.

In this work, we investigate whether this issue can be overcome by considering non-thermal production of the mother particle. Specifically, we consider production via the decay of an exotic matter field dominating the energy budget of the Universe after inflation. As a proof-of-principle, we apply this idea to a simplified model, i.e. an extension of the SM with the minimal field content needed to accommodate both baryogenesis and DM production, while remaining agnostic on an eventual UV completion. The effects of non-standard cosmological history are parametrized via: i) the so-called reheating temperature, which represents the temperature at which the Universe enters the radiation-dominated epoch % returns to be radiation-dominated
after the decay of the exotic matter component, and ii) the decay branching fractions of the latter into radiation and into the particles responsible for the generation of the baryon asymmetry and the DM abundance.

The paper is structured as follows. Section~\ref{sec:simplified_model} and Section~\ref{sec:non-thermal_baryogen} are mostly devoted to the generation of the baryon asymmetry. More specifically, we introduce in Section~\ref{sec:simplified_model} the simplified model first presented by ~\cite{Claudson:1983js}, and briefly review how baryogenesis is accommodated assuming a standard thermal history for the Universe. In Section~\ref{sec:non-thermal_baryogen}, we instead illustrate the non-standard cosmological scenario and present the main numerical results. In Section~\ref{sec:including_DM}, we finally illustrate some possibilities to incorporate a DM candidate. In this last case we limit our analysis to some benchmarks, leaving a more systematic study to the future. We then present our conclusions in Section~\ref{sec:conclusions}.

\section{Simplified model for the production of the baryon asymmetry}
\label{sec:simplified_model}
The minimal field content to accommodate the generation of a non-zero $\CP$-asymmetry is represented by two massive Majorana fermions, denoted $X_r$ ($r=1,2$) and a scalar field $\Phi$ interacting via a Yukawa-like Lagrangian reading as~\cite{Claudson:1983js}:
    \begin{equation}
    \label{eq:model_lagrangian}
        \mathcal{L}=A_{ra} X_r D_a \Phi+B_{ab}U_a D_b \Phi^{*}+\frac{1}{2}C_{ab}Q_a Q_b \Phi\,.
    \end{equation}
Here, $Q$ and $U,D$ stand, respectively, for the $SU(2)$ singlet and doublet quark fields, with $a,\,b$ generation indices. 
For definiteness, we assume that $M_\Phi> M_{X_{2}} > M_{X_{1}}$ \footnote{The case $M_\Phi < M_{X_2}$ has been considered, for example, in \cite{Cui:2012jh}}. In addition to the interactions of Eq.~\eqref{eq:model_lagrangian}, responsible for the generation of the baryon asymmetry, we assume that the state $\Phi$ features B-conserving interactions (e.g., gauge interactions), ensuring it is always in thermal equilibrium.
If a standard cosmological history is assumed, the generation of the baryon asymmetry can be determined via a system of coupled Boltzmann's equations tracking the number density of the $X_{r=1,2}$ fields, dubbed $n_{X_r}$, as well as the difference $B-L$ of the baryon and lepton numbers\footnote{We track $B-L$ in order to avoid including explicitly the rate of the sphaleron processes. In fact, the latter might wash away the generated baryon asymmetry, unless this has a non-zero projection on $B-L$. See, e.g., Ref.~\cite{Kolb:1990vq}}, $n_{B-L}$. The most general form of such a system reads:
%%%%%%%%%%%%%%%%%%%%%%%%%%%%%%%%%%%%%%%%%%
    \begin{align}
    \label{eq:boltzmann_simplified_model}
    \begin{split}
        \frac{dn_{X_1}}{dt}+3Hn_{X_1}&=- \left[\langle \Gamma_1 \rangle +\langle \sigma_{1q} v \rangle n_{q,\eq}\right]\left(n_{X_1}-n_{X_1,\eq}\right)\\
        &+\left[\langle \Gamma_{21} \rangle +\langle \sigma_{Xq} v \rangle n_{q,\eq}\right]\left(n_{X_2}-\frac{n_{X_2,\eq}}{n_{X_1,\eq}}n_{X_1}\right)\\
        & -\langle \sigma_{12} v \rangle \left(n_{X_1}n_{X_2}-n_{X_1,\eq}n_{X_2,\eq}\right)-2 \langle \sigma_1 v \rangle  \left(n_{X_1}^2-n_{X_1,\eq}^2\right)\\
        \frac{dn_{X_2}}{dt}+3Hn_{X_2}&=-\left[\langle \Gamma_2 \rangle +\langle \sigma_{2q} v \rangle n_{q,\eq}\right]\left(n_{X_2}-n_{X_2,\eq}\right)\\
        &-\left[\langle \Gamma_{21} \rangle +\langle \sigma_{Xq} v \rangle n_{q,\eq}\right]\left(n_{X_2}-\frac{n_{X_2,\eq}}{n_{X_1,\eq}}n_{X_1}\right)\\
        & -\langle \sigma_{12} v \rangle \left(n_{X_1}n_{X_2}-n_{X_1,\eq}n_{X_2,\eq}\right)-2 \langle \sigma_2 v \rangle  \left(n_{X_2}^2-n_{X_2,\eq}^2\right)\\
        \frac{dn_{B-L}}{dt}+3Hn_{B-L}&=\langle \Delta \Gamma \rangle \left(n_{X_2}-n_{X_2,\eq}\right)+\langle \Delta \sigma v \rangle n_{q,\eq}\left(n_{X_2}-\frac{n_{X_2,\eq}}{n_{X_1,\eq}}n_{X_1}\right)\\
        & -\frac{3}{2}\frac{n_{B-L}}{n_{q,\eq}}\sum_{r=1,2}\big[3 \langle \Gamma_r \rangle n_{q,\eq}+\langle \sigma_{rq} v \rangle n_{q,\eq}\left(n_{X_r}+2 n_{X_r,\eq}\right)\big]
        % &\frac{dn_1}{dt}+3Hn_1=- \left[\langle \Gamma_1 \rangle +\langle \sigma_{1q} v \rangle n_{q,\eq}\right]\left(n_1-n_{1,\eq}\right) +\left[\langle \Gamma_{21} \rangle +\langle \sigma_{Xq} v \rangle n_{q,\eq}\right]\left(n_2-\frac{n_{2,\eq}}{n_{1,\eq}}n_1\right)\\
        % & -\langle \sigma_{12} v \rangle \left(n_1n_2-n_{1,\eq}n_{2,\eq}\right)-2 \langle \sigma_1 v \rangle  \left(n_1^2-n_{1,\eq}^2\right)\\
        % %%
        % &\frac{dn_2}{dt}+3Hn_2=-\left[\langle \Gamma_2 \rangle +\langle \sigma_{2q} v \rangle n_{q,\eq}\right]\left(n_2-n_{2,\eq}\right)-\left[\langle \Gamma_{21} \rangle +\langle \sigma_{Xq} v \rangle n_{q,\eq}\right]\left(n_2-\frac{n_{2,\eq}}{n_{1,\eq}}n_1\right)\\
        % & -\langle \sigma_{12} v \rangle \left(n_1n_2-n_{1,\eq}n_{2,\eq}\right)-2 \langle \sigma_2 v \rangle  \left(n_2^2-n_{2,\eq}^2\right)\\
        % %%
        % &\frac{dn_{B-L}}{dt}+3Hn_{B-L}=\langle \Delta \Gamma \rangle \left(n_2-n_{2,\eq}\right)+\langle \Delta \sigma v \rangle n_{q,\eq}\left(n_2-\frac{n_{2,\eq}}{n_{1,\eq}}n_1\right)\\
        % & -\frac{3}{2}\frac{n_{B-L}}{n_{q,\eq}}\sum_{r=1,2}\big[3 \langle \Gamma_r \rangle n_{q,\eq}+\langle \sigma_{rq} v \rangle n_{q,\eq}\left(n_{r}+2 n_{r,\eq}\right)\big]
    \end{split}
    \end{align}
Here, $H=\dot{a}/a$ is the Hubble expansion rate, with $a$ the cosmological expansion parameter and the dot representing time derivative. The subscript ``eq'' indicates equilibrium densities, defined as~\cite{Claudson:1983js}
    \begin{align*}
        n_{r,\eq} = \frac{M_{X_r}^2T}{\pi^2}K_2\left(\frac{M_{X_r}}{T}\right)\,,\quad
        n_{q,\eq} = \frac{36T^3}{\pi^2}\,,
    \end{align*}
where $K_n(x)$ denotes the modified Bessel function of order $n$. Lastly, thermal averages are represented with angle brackets. 

In the first two equations of \eqref{eq:boltzmann_simplified_model}, the first line describes the $B$-violating decay and annihilation of the single states, the second line describes decay and scattering involving both beyond-the-Standard-Model (BSM) states and the third line describes co-annihilation and pair annihilation. The cross-sections of the different processes can be written as~\cite{Claudson:1983js}:
\begin{align} 
\label{eq:ann_processes}
%\begin{split}
& \langle\sigma_{r}v\rangle \equiv \langle \sigma_{X_r X_r \rightarrow q \bar q} v \rangle=\frac{27 \lambda^4}{128\pi}\frac{M_{X_r}^2}{M_\Phi^4}\left[ \left(\frac{K_3\left(M_{X_r}/T\right)}{K_2\left(M_{X_r}/T\right)}\right)^2-\left(\frac{K_1\left(M_{X_r}/T\right)}{K_2\left(M_{X_r}/T\right)}\right)^2\right],\nonumber\\
& \langle\sigma_{12}v\rangle \equiv \langle \sigma_{X_1 X_2 \rightarrow q \bar q} v  \rangle=\frac{9 \lambda^4}{64\pi}\frac{M_{X_2}^2}{M_\Phi^4}\left[2 \frac{K_4\left(M_{X_2}/T\right)}{K_2\left(M_{X_2}/T\right)}+1\right]\,,\nonumber\\ % B-conserving
&\langle\sigma_{Xq}v\rangle \equiv \langle \sigma_{X_2 q \rightarrow X_1 q} v \rangle=\frac{9 M_{X_2}^2}{128\pi M_\Phi^4}\left[8 \frac{K_4\left(M_{X_2}/T\right)}{K_2\left(M_{X_2}/T\right)}+1\right],\nonumber\\ % B-conserving
& \langle\sigma_{rq}v\rangle \equiv\langle \sigma_{X_r \bar q \rightarrow qq} v + \sigma_{X_r q \rightarrow \bar q\bar q} v \rangle =\frac{117 \lambda^4 M_{X_r}^2}{576\pi M_\Phi^4}\left[5 \frac{K_4\left(M_{X_r}/T\right)}{K_2\left(M_{X_r}/T\right)}+1\right], % B-violating
%    \end{split}
\end{align}
under the assumption that all the couplings in Eq.~\eqref{eq:model_lagrangian} are equal to the same value $\lambda$. All the rates (annihilations and decays) presented in this work are summed over all the quark flavors in the final state. 
Both ${X_1}$ and ${X_2}$ are unstable and decay with the following rates:
\begin{equation}
\label{eq:GammaBp}
        \Gamma_{21} \equiv \Gamma_{X_2 \rightarrow X_{1} q \bar q =} \frac{9 \lambda^4  M_{X_2}^5}{1024\pi^3 M_\Phi^4},
    \end{equation}
  \begin{equation}
  \label{eq:GammaBv}
        \Gamma_r \equiv \Gamma_{X_r \rightarrow qqq}+\Gamma_{X_r \rightarrow \bar q \bar q \bar q}=\frac{117 M_{X_r}^5}{1024\pi^3 M_\Phi^4}\,,
    \end{equation}   
whose thermal averages are computed with the customary formula~\cite{Claudson:1983js}:
\begin{equation}
        \langle \Gamma_{A\to B} \rangle=\Gamma_{A\to B} \frac{K_1\left(M_{A}/T\right)}{K_2\left(M_{A}/T\right)}\,.
    \end{equation}
As already discussed in Ref.~\cite{Arcadi:2015ffa} (see also Ref.~\cite{Rompineve:2013grm}), the simultaneous presence of B-violating and B-preserving decay channels for the heaviest particle $X_2$ is a mandatory requirement to a have a non-zero asymmetry. 

The third equation of \eqref{eq:boltzmann_simplified_model} tracks the evolution of $B-L$. The first line contains the sources of $\CP$-asymmetry. This is generated both from decays, governed by~\cite{Claudson:1983js}
    \begin{align}
        \Delta \Gamma &\equiv\Gamma_{X_2 \rightarrow qqq}-\Gamma_{X_2 \rightarrow \bar q \bar q \bar q} =\frac{9 \lambda^6 M_{X_2}^7}{40960\pi^4 M_\Phi^6}
    \end{align}
and single annihilation processes, described by
%\AM{What is the summation over? Edit: Removed it, see \cite{Claudson:1983js} appendix}
%
    \begin{align}
    \begin{split}
        \langle \Delta \sigma v \rangle &\equiv \langle \sigma_{X_2 \bar q \rightarrow qq} v \rangle -\langle \sigma_{X_2 q \rightarrow \bar q \bar q} v \rangle = \frac{3 %9
        \lambda^6 M_{X_2}^4}{1024 %3072
        \pi^2 M_\Phi^6}%\sum_{abcd}
        \Big[\frac{K_4\left(M_{X_2}/T\right)}{K_2\left(M_{X_2}/T\right)}
        +\frac{6T}{M_{X{2}}}\frac{K_5\left(M_{X_2}/T\right)}{K_2\left(M_{X_2}/T\right)}\Big
        ]\,.
    \end{split}
    \end{align}

In order to numerically solve the system~\eqref{eq:boltzmann_simplified_model}, it is more convenient to work with dimensionless variables: $n_{X_i} \rightarrow Y_{X_i}=n_{X_i}/s$, $t \rightarrow x=M_{X_2}/T$. Here, $s = \frac{2\pi^2}{45}h_\mathrm{eff} T^3$ is the entropy density and $h_\mathrm{eff}$ is the effective number of entropy degrees of freedom. Then, the system of Boltzmann's equations reads
\begin{equation}
\begin{aligned}
%\begin{split}
\label{eq:dimensionless_boltzmann_simplified_model}
     \frac{dY_{X_1}}{dx} &= \frac{1}{Hx}\Big[\Big.-\langle \Gamma_1 \rangle  \left(Y_{X_1}-Y_{X_1,\rm eq}\right)-s\langle \sigma_{1q} v \rangle Y_{q,\eq}\left(Y_{X_1}-Y_{X_1,\eq}\right)\\
     &+\langle \Gamma_{21} \rangle \left(Y_{X_2}-\frac{Y_{X_2,\eq}}{Y_{X_1,\eq}}Y_{X_1}\right)+s\langle \sigma_{Xq} v \rangle  Y_{q,\eq} \left(Y_{X_2}-\frac{Y_{X_2,\eq}}{Y_{X_1,\eq}}Y_{X_1}\right)\\
     &-s\langle \sigma_{12} v \rangle \left(Y_{X_1} Y_{X_2}-Y_{X_1,\eq}Y_{X_2,\eq}\right) -2 s \langle \sigma_1 v \rangle \left(Y_{X_1}^2-Y_{X_1,\eq}^2\right)\Big.\Big]\\
    % \end{aligned}
    % \end{equation*}
    % \begin{equation*}
    % \begin{aligned}
    \frac{dY_{X_2}}{dx} &= \frac{1}{Hx}\Big[\Big.-\langle \Gamma_2 \rangle \left(Y_{X_2}-Y_{X_2,\rm eq}\right)-s\langle \sigma_{2q} v \rangle  Y_{q,\eq}\left(Y_{X_2}-Y_{X_2,\eq}\right)\\
    &-\langle \Gamma_{21} \rangle \left(Y_{X_2}-\frac{Y_{X_2,\eq}}{Y_{X_1,\eq}}Y_{X_1}\right) -s\langle \sigma_{Xq} v \rangle  Y_{q,\eq} \left(Y_{X_2}-\frac{Y_{X_2,\eq}}{Y_{X_1,\eq}}Y_{X_1}\right) \\
    &-s\langle \sigma_{12} v \rangle \left(Y_{X_1} Y_{X_2}-Y_{X_1,\eq}Y_{X_2,\eq}\right)-2 s  \langle \sigma_2 v \rangle \left(Y_{X_2}^2-Y_{X_2,\eq}^2\right)\Big.\Big] \\
    % \end{aligned}
    % \end{equation*}
    % \begin{equation}
    % \begin{aligned}
    \frac{dY_{B-L}}{dx} &= \frac{1}{Hx}\Big[\Big.\langle \Delta \Gamma \rangle \left(Y_{X_2}-Y_{X_2,\eq}\right)+s\langle \Delta \sigma v \rangle Y_{q,\eq}\left(Y_{X_2}-\frac{Y_{X_2,\eq}}{Y_{X_1,\eq}}Y_{X_1}\right)\\
    & -\frac{3}{2}\frac{Y_{B-L}}{Y_{q,\eq}}\sum_{r=1,2}\big(3 \langle \Gamma_r \rangle Y_{X_r,\eq}+\langle \sigma_{rq} v \rangle s Y_{q,\eq}\left(Y_{X_r}+2 Y_{X_r,\eq}\right)\big)\Big.\Big]
%\end{split}
\end{aligned}
\end{equation}
%%%%%%%%%%%%%%%%%%%%%%%%%%%%%%%%%%%%%%%%%%%
% \begin{figure*}
%     \centering        \subfloat{\includegraphics[width=0.35\linewidth]{figures_draft/pbensimp.pdf}}\\    \subfloat{\includegraphics[width=0.33\textwidth]{figures_draft/pben1th.pdf}}    \subfloat{\includegraphics[width=0.33\textwidth]{figures_draft/pben2th.pdf}}    \subfloat{\includegraphics[width=0.33\textwidth]{figures_draft/pben3th.pdf}}
%     \caption{\footnotesize{ {\it Upper row:} $Y_{B-L}$ as function of the coupling $\lambda$ for three benchmark assignations of the $(M_{X_1},M_{X_2},M_\Phi)$ set, reported on the plot.{\it Lower row:} Evolution of $Y_{X_{1,2}},Y_{B-L}$ as function of $x=M_{X_2}/T$, assuming a standard cosmological history, for the same benchmarks considered in the top row. The value of the couplings correspond to the one providing, in each case, the maximal value of $Y_{B-L}$.}}
%     \label{fig:pbenth}
% \end{figure*}

We have numerically solved the system~\eqref{eq:dimensionless_boltzmann_simplified_model}, with initial conditions at $x \ll 1$\footnote{For definiteness we have considered $x=10^{-3}$ as initial condition but we have verified that the solution is independent on the initial value of $x$, as long as $x \ll 1$.}, $Y_{X_1}=Y_{X_2}=Y_{B-L}=0$, for three benchmark assignations for the set $\left( M_{X_{1}},M_{X_{2}},M_\Phi\right)$. %, namely $\left(80\,\mbox{GeV},200\,\mbox{GeV},3\times 10^5\,\mbox{GeV}\right)$, $\left(4\,\mbox{TeV},10\,\mbox{TeV},10^6\,\mbox{GeV}\right)$, $\left(10\,\mbox{TeV},30\,\mbox{TeV},10^6\,\mbox{GeV}\right)$.\AM{What is the rationale behind these choices? It seems none of these choices reproduces the observed yield. Do we want to say something like ``since this model tends to underproduce B-L, we introduce a new mechanism that enhances it''?} 

\begin{figure}
    \centering        \subfloat{\includegraphics[width=0.4\linewidth]{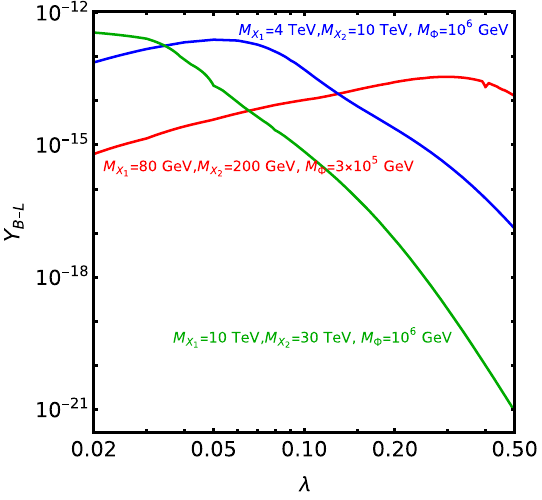}}
    \caption{\footnotesize{Final ($x\gg1$) value of $Y_{B-L}$, obtained by numerical integration of the system~\eqref{eq:dimensionless_boltzmann_simplified_model}, as function of the coupling $\lambda$ and for three benchmark assignations of the $(M_{X_1},M_{X_2},M_\Phi)$ set, reported as colored labels on the plot.}}
    \label{fig:yield_simplified_model}
\end{figure}
\begin{figure}
    \centering       \subfloat{\includegraphics[width=0.33\textwidth]{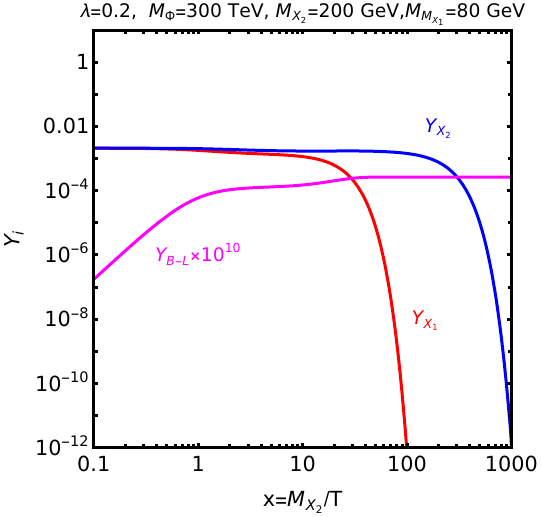}}    \subfloat{\includegraphics[width=0.33\textwidth]{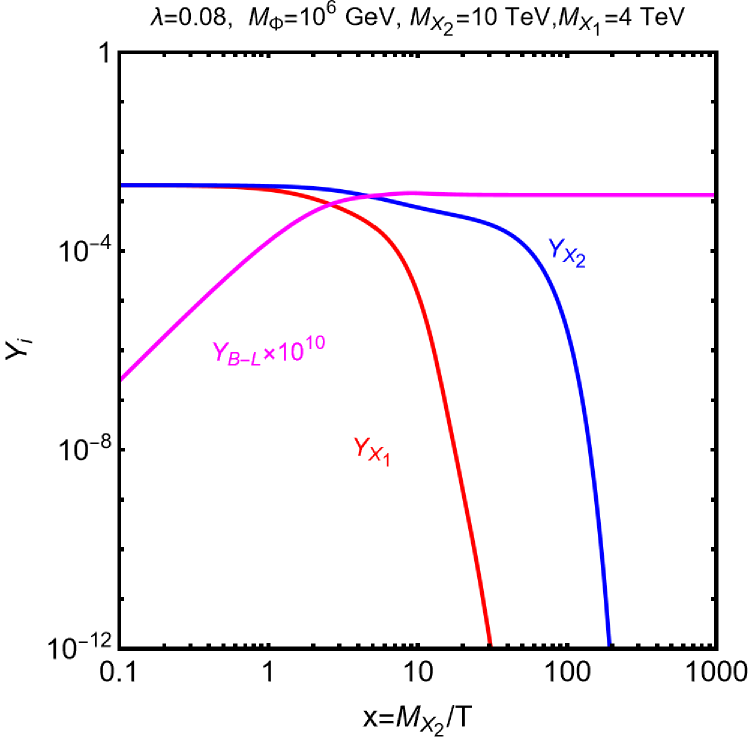}}    \subfloat{\includegraphics[width=0.33\textwidth]{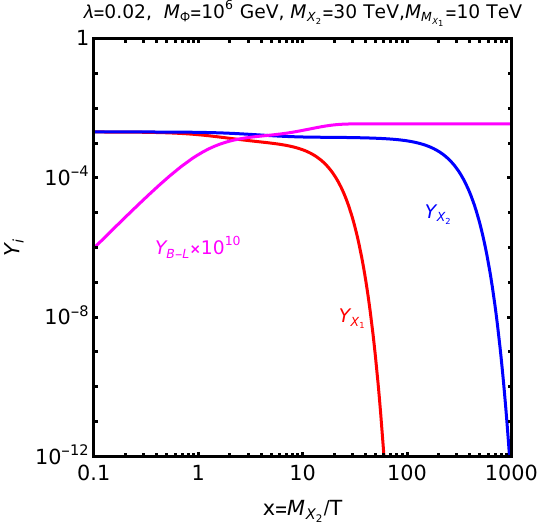}}
    \caption{\footnotesize{Evolution of $Y_{X_{1}}$ (red), $Y_{X_{2}}$ (blue) and $Y_{B-L}$ (magenta) as function of the time parameter $x=M_{X_2}/T$, assuming a standard cosmological history (system~\eqref{eq:dimensionless_boltzmann_simplified_model}), for the same benchmarks considered in Fig.\ref{fig:yield_simplified_model} and reported on top of each plot. The value of the couplings correspond to the one providing, in each case, the maximal value of $Y_{B-L}$. (see Fig.\ref{fig:yield_simplified_model}).}}
    \label{fig:evolution_simplified_model}
\end{figure}
In Fig.~\ref{fig:yield_simplified_model} we show the final yield of the $B-L$ asymmetry, i.e. the value of $Y_{B-L}$ for $x\gg1$, as a function of the coupling $\lambda$ for the three considered benchmarks, with values specified in the colored labels. The behavior of $Y_{B-L}$ is consistent across the benchmarks, increasing with the value of $\lambda$, reaching a maximum and then decreasing as $\lambda$ approaches its maximum value $\lambda=1$. This can be explained in terms of two competing effects: on the one hand, the $B-L$ asymmetry is proportional to the $\CP$ asymmetry in the decay and annihilation processes of the $X_2$ particles, which increases with the value of $\lambda$; on the other hand, a higher value of $\lambda$ corresponds to more efficient annihilations of $X_2$ and hence to a lower abundance, which is in turn translated to a lower $B-L$ abundance. As found e.g. in~\cite{Arcadi:2015ffa}, a similar argument hold for the ratio $M_{X_2}/M_\Phi$. A low value would correspond to high $CP$ asymmetry. At the same time, however, the decay and annihilation rates of $X_2$ would be enhanced, as well as possibly the wash-out rates related to $X_1$, corresponding to a low abundance of the mother particle and, consequently, a lower baryon abundance. % The choices of the set $(M_{X_1},M_{X_2},M_\Phi)$ for the benchmarks shown in fig. (\ref{fig:yield_simplified_model})  already stated

Fig.~\ref{fig:evolution_simplified_model} shows the evolution of the abundances of the $X_{1}$ and $X_{2}$ states and of $B-L$, with respect to the time variable $x=M_{X_2}/T$. In all cases we have considered the value of $\lambda$ which yields the highest value of $Y_{B-L}$ (see Fig.~\ref{fig:yield_simplified_model}). We notice that, for all benchmark assignations, the particles $X_{1}$ and $X_{2}$ encounter an early decoupling, with respect to the typical WIMP case, very close to the relativistic regime followed by their decay. On the other hand, $Y_{B-L}$ features a rather sharp increase and reaches a first plateu in correspondence of the decoupling of $X_1$ from the thermal plasma. This indicates a sizable contribution to $Y_{B-L}$ from $\CP$- and $B$- violating annihilation processes of $X_1$. For the benchmarks in the first and third panel of Fig.~\ref{fig:evolution_simplified_model}, $Y_{B-L}$ increases again at later times and reaches its final value in correspondence of the decay of $X_{2}$, indicating an additional contribution from $\CP$- and $B$-violating decays.

Neither of the three benchmarks is able to produce an amount  of $Y_{B-L}$ corresponding to the experimentally favored value $Y_{B-L}^{\rm obs}\simeq10^{-10}$. As already pointed out, thermal production can hardly account for a sufficient abundance of the mother particle to compensate the small value of the $\CP$ asymmetry. This motivates us to explore the scenario of non-thermal production and/or non-standard cosmological histories.

\section{Baryogenesis from non-thermal production}
\label{sec:non-thermal_baryogen}
We now consider a non-standard cosmological scenario dubbed Early Matter Domination (EMD)~\cite{Arcadi:2015ffa}. It consists in the introduction of an additional metastable matter component, described by a field $\Psi$, decoupled from the primordial plasma and dominating the energy budget of the Universe for a certain stage between primordial inflation and %the temperature of onset of the
Big Bang Nucleosynthesis (BBN). The additional matter component must decay before BBN, not to spoil the successful Standard Cosmological model prediction of the light nuclei abundances. Specifically, $\Psi$ can decay into (at least) $X_2$ particles, non-thermally enhancing their abundance and, in turn, the baryon abundance. 
Following these prescriptions, we modify the system~\eqref{eq:boltzmann_simplified_model}
% \ref{eq:base_system}
by adding two equations which trace the evolution of the energy density of the $\Psi$ field and of the thermal plasma~\cite{Chung:1998rq}:
%\AM{I think the term $\langle E_{X_r}\rangle$ in the second equation should be outside of the square bracket. SK:YES, Eq. 3 of https://arxiv.org/pdf/hep-ph/9809453 }
% Having these considerations in mind, we complement the system (\ref{eq:boltzmann_simplified_model}) with two further equations:
\begin{align}
\begin{split}
\label{eq:Boltzmann_add}
    & \frac{d\rho_\Psi}{dt}+3H \rho_\Psi=-\Gamma_\Psi \rho_\Psi\,,\\
    & \frac{d\rho_R}{dt}+4H\rho_R=\left(1-\frac{\sum_{r=1,2} B_{X_r}\langle E_{X_r} \rangle}{M_\Psi}\right)\Gamma_\Psi \rho_\Psi\\
    &\quad+ \sum_{r=1,2}\langle E_{X_r}\rangle \Bigg[ \Gamma_{X_r} n_{X_r}+ \langle \sigma_r v \rangle \left(n_{X_r}^2-n_{X_r,\eq}^2\right) + \langle \sigma_{rq} v \rangle n_{q,\eq}\left(n_{X_r}-n_{{X_r,\eq}}\right)\Big] \\
    &\quad+ \langle E_{X_2} \rangle \Bigg[\langle\sigma_{Xq} v \rangle\left(n_{X_2}-\frac{n_{X_1}}{n_{{X_1,\eq}}}n_{X_{2,\eq}}\right)+\langle \sigma_{12} v \rangle\left(n_{X_1} n_{X_2}-n_{{X_1,\eq}} n_{{X_2,\eq}}\right)\Bigg]\,.
   \end{split}
\end{align} 
Here, $\langle E_{X_r} \rangle=\sqrt{M_{X_r}^2+3 T^2}$, while $B_{X_r}$ represent the decay branching fractions of $\Psi$ into $X_r$, times the multiplicity of particles produced in each decay. The Hubble expansion rate is given by:
\begin{equation}
\label{eq:full_Hubble}    H=\sqrt{\frac{\rho_R+\rho_\Psi+\sum_{r=1,2} \langle E_{X_{r}} \rangle n_{X_r}}{3 M_{\rm Pl}^2}}
\end{equation}
with $\rho_R \propto T^4$. As already pointed out, we assume suitable initial conditions for $\rho_\Psi$ so that it dominates the energy budget of the Universe between the freeze-out of the $X_2$ state and the time of onset of the BBN. For completeness Eq.~\eqref{eq:full_Hubble} includes also contributions from the energy densities of the $X_{r}$ states, potentially sizable in presence of very efficient non-thermal production.
% \AM{Maybe move below. Edit: moved below} During the EMD, the relation between the temperature and the scale factor becomes $T \propto a^{-3/8}$~\cite{Giudice:2000ex}, contrary to the conventional scaling $T \propto a^{-1}$ during the standard radiation domination.

The equation for the $\Psi$ state, in \eqref{eq:Boltzmann_add}, is rather simple as the time evolution of its energy density depends on the expansion rate of the Universe and the decay rate $\Gamma_\Psi$ of the $\Psi$ state itself.
Remaining agnostic on the specific form of the interaction between $\Psi$ and the SM, we parameterize its decay rate in terms of the reheating temperature $T_{R}$, i.e. the temperature at which the Universe becomes radiation-dominated after the decay of $\Psi$. Adopting the customary instantaneous reheating approximation, we can express% the decay rate $\Gamma_\Psi$ in terms of the Hubble rate at $T_R$:
\begin{equation}    \Gamma_\Psi=H(T_R)=\sqrt{\frac{\pi^2g_\mathrm{eff}(T_R)}{90M_{Pl}^2}}T_R^2
\end{equation}
where $g_\mathrm{eff}(T)$ is the effective number of degrees of freedom contributing to the radiation energy density. 
The requirement that $\Psi$ decays before the onset of BBN translates into the lower bound $T_R \gtrsim 5\,\mbox{MeV}$~\cite{Barbieri:2025moq}. Moving to the equation for the radiation energy density, we point out that it can receive a sizable contribution from the decay of $\Psi$ and from both decay and annihilation processes into radiation of the $X_i$ particles, which might in turn dominate the energy density of the Universe in case of very efficient non-thermal production. 
%\AM{Redundant with a footnote} 
While such contributions have been reported, for completeness, in Eqs.~\eqref{eq:Boltzmann_add}, we have verified, at least for the benchmarks considered in present study, that their contribution to $\rho_R$ is actually negligible. 

While we regard the masses $M_{X_r}$ and the branching fractions $B_{X_r}$ as free parameters, for definiteness we assume a fixed value $M_{\Psi}=10^6\,\mbox{GeV}$ for the mass of the exotic matter field $\Psi$. 

Following analogous steps as in, for example~\cite{Giudice:2000ex,Arcadi:2011ev}, we introduce the dimensionless variables:
    \begin{align}
        \xi_{\Psi}&=\rho_\Psi \Lambda^{-1} a^3,\\
        \quad N_{X_r}&=n_{X_r} a^3,\\
        \quad N_{B-L}&=n_{B-L}a^3,
    \end{align}
where $\Lambda$ is an arbitrary energy scale. 
Furthermore, the equation for $\rho_R$ can be converted into an equation tracking the temperature of the primordial plasma~\cite{Giudice:2000ex,Arcadi:2011ev}:
    \begin{align}
        \rho_R(T)=\frac{3}{4}\frac{g_\mathrm{eff}(T)}{h_\mathrm{eff}(T)}Ts(T)\,.
    \end{align}
Lastly, we use as time variable the dimensionless scale factor $A=a \Lambda$.
Combining Eq.s~\eqref{eq:boltzmann_simplified_model} with Eq.s~\eqref{eq:Boltzmann_add} and working with the new variables, we obtain:
%%%%%%%%%%%%%%%%%%%%%%%%%%%%%%%%%%%%%%%%%%%
    \begin{align}
    \begin{split}
    \label{eq:BoltzNT_Psi}
        \frac{d\xi_\Psi}{dA}&=-\Gamma_\Psi\xi_\Psi\frac{A^{1/2}\Lambda^{-3/2}}{\mathcal{H}}\,,
    \end{split}
    \end{align}
    \begin{align}
    \begin{split}
    \label{eq:BoltzNT_T}
        \frac{dT}{dA}&=\left(1+\frac{T}{4 g_{\rm eff}}\frac{dg_{\rm eff}}{dT}\right)^{-1}\Bigg\{-\frac{h_{\rm eff}}{g_{\rm eff}}\frac{T}{A}+\frac{h_\mathrm{eff}}{3g_{\rm eff}s(T)}\frac{A^{-5/2}\Lambda^{3/2}}{\mathcal{H}}\Bigg[\left(1-\sum_{r=1,2}\frac{\langle E_{X_r}\rangle}{M_\Psi}B_{X_r}\right)\Lambda \Gamma_\Psi \xi_\Psi \Bigg.\\
        &+ \sum_{r=1,2}\frac{\langle E_{r} \rangle \Lambda^3}{A^3}\bigg[\langle \sigma_r v \rangle \left(N_{X_r}^2-N_{X_r,\eq}^2\right) + \langle \sigma_{rq} v \rangle N_{q,\eq}\left(N_{X_r}-N_{{X_r,\eq}}\right)\\
        &+ \Bigg. \Bigg. \bigg. \langle \sigma_{Xq} v \rangle\left(N_{X_2}-\frac{N_{X_1}}{N_{{X_1,\eq}}}N_{X_{2,\eq}}\right)+\langle \sigma_{12} v \rangle\left(N_{X_1} N_{X_2}-N_{{X_1,\eq}} N_{{X_2,\eq}}\right)\bigg]\Bigg] \Bigg\}, 
    \end{split}
    \end{align}
    \begin{align}
   \begin{split}
    \label{eq:BoltzNT_N1}
        \frac{dN_{X_1}}{dA}&=-\frac{A^{-5/2}\Lambda^{3/2}}{\mathcal{H}}\bigg[\langle \sigma_1 v \rangle \left(N_{X_1}^2-N_{X_1,\eq}^2\right)+ \langle \sigma_{1q} v \rangle N_{q,\eq}\left(N_{X_1}-N_{X_1,\eq}\right)\\
        &-\langle \sigma_{Xq} v \rangle\left(N_{X_2}-\frac{N_{X_1}}{N_{X_1,\eq}}N_{X_2,\eq}\right) +\langle \sigma_{12} v \rangle \left(N_{X_1} N_{X_2}-N_{X_1,\eq} N_{X_2,\eq}\right)\bigg]\\
        &-\frac{A^{1/2}\Lambda^{-3/2}}{\mathcal{H}}\Big[\langle \Gamma_1 \rangle \left(N_{X_1}-N_{X_1,\eq}\right)-\langle \Gamma_{21} \rangle \left(N_{X_2}-\frac{N_{X_1}}{N_{X_1,\eq}}N_{X_2,\eq}\right) \Big] \\
        &+ \Lambda\frac{B_{X_1}}{M_\Psi}\Gamma_\Psi \xi_\Psi \frac{A^{1/2}\Lambda^{-3/2}}{\mathcal{H}}\,,
    \end{split}
    \end{align}
    \begin{align}
   \begin{split}
    \label{eq:BoltzNT_N2}
        \frac{dN_{X_2}}{dA}&=-\frac{A^{-5/2}\Lambda^{3/2}}{\mathcal{H}} \bigg[\langle \sigma_2 v \rangle \left(N_{X_2}^2-N_{X_2,\eq}^2\right)+ \langle \sigma_{2q} v \rangle N_{q,\eq}\left(N_{X_2}-N_{X_2,\eq}\right)\\
        &-\langle \sigma_{Xq} v \rangle\left(N_{X_2}-\frac{N_{X_1}}{N_{X_1,\eq}}N_{X_2,\eq}\right) +\langle \sigma_{12} v \rangle \left(N_{X_1} N_{X_2}-N_{X_1,\eq} N_{X_2,\eq}\right)\bigg]\\
        &-\frac{A^{1/2}\Lambda^{-3/2}}{\mathcal{H}}\Big[\langle \Gamma_2 \rangle \left(N_{X_2}-N_{X_2,\eq}\right) -\langle \Gamma_{21} \rangle \left(N_{X_2}-\frac{N_{X_1}}{N_{X_1,\eq}}N_{X_2,\eq}\right) \Big]\\
        & + \Lambda\frac{B_{X_2}}{M_\Psi}\Gamma_\Psi \xi_\Psi \frac{A^{1/2}\Lambda^{-3/2}}{\mathcal{H}}\,,
   \end{split}
    \end{align}
    \begin{align}
   \begin{split}
    \label{eq:BoltzNT_BL}
        \frac{dN_{B-L}}{dA}&=\frac{A^{1/2}\Lambda^{-3/2}}{\mathcal{H}}\langle \Delta \Gamma \rangle\left(N_{X_2}-N_{X_2,\rm eq}\right)+\frac{A^{-5/2}\Lambda^{3/2}}{\mathcal{H}}\langle \Delta \sigma v \rangle N_{q,\eq} \left(N_{X_2}-N_{X_2,\rm eq}\right)\\
        & -\sum_{r=1,2}\frac{9}{2}\frac{A^{1/2}\Lambda^{-3/2}}{\mathcal{H}}\langle\Gamma_r \rangle  N_{{X_r,\rm eq}}-\sum_{r=1,2}\frac{3}{2}\frac{A^{-5/2}\Lambda^{  3/2}}{\mathcal{H}}\langle \sigma_r v \rangle N_{B-L}\left(N_{X_r}+2N_{{X_r,\rm eq}}\right).
   \end{split}
    \end{align}
The factor $\mathcal{H}$ is related to the Hubble parameter $H$ by:
    \begin{align}
        & \mathcal{H}=(A/\Lambda)^{3/2}H={\left(\frac{\Lambda \xi_\phi+\rho_R (T) (A/\Lambda)^3+\sum_{r=1,2}  E_{X_r} N_{X_r}}{3 M_{\rm Pl}^2}\right)}^{1/2}
    \end{align}
% The differential equations are solved in terms of the adimensional scale factor $A=a \Lambda$.
% the equation for $\rho_R$ is converted into an equation tracking the temperature of the primordial plasma.
In Eq.~\eqref{eq:BoltzNT_T}, tracking the temperature of the primordial plasma, the first term in curly brackets represents the standard cosmological contribution: if only this term was present, we would recover the solution $T \propto a^{-1}$ typical of a radiation-dominated Universe. During the EMD, the relation between the temperature and the scale factor becomes instead $T \propto a^{-3/8}$~\cite{Giudice:2000ex}. %, contrary to the conventional scaling $T \propto a^{-1}$ during the standard radiation domination. %\textcolor{red}{Q: This relation depends on entropy conservation, so hold througout evolution in standard scenario irrespective of radiation era??}. 
The term proportional to $\Gamma_\Psi$ in the square brackets describes the entropy injection due to the decay of the scalar field. The remaining terms account for analogous effects from decays and annihilation of the $X_{1}$ and $X_{2}$ states.
%~\footnote{These contributions have been included in the Boltzmann's equations for completeness. We have nevertheless verified that, at least for the benchmarks considered in our study, their impact is negligible.}.
Finally, we follow analogous prescriptions as in Ref.~\cite{Giudice:2000ex,Arcadi:2011ev}, setting the initial energy density of $\Psi$ to be $\rho_{\Psi,I}=\frac{1}{2}M_\psi^2 M_{\rm Pl}^2$ while $X_{1,I}=X_{2,I}=X_{B-L,I}=0$. The initial temperature has been set to $10^6 \,\mbox{GeV}$ (we have verified that the solution of the Boltzmann's equation is not affected by changing the initial value of the temperature).
Furthermore, we set the branching fraction of the lighter Majorana particle, $B_{X_1}=0$.
\begin{figure}
    \centering
    \subfloat{\includegraphics[width=0.5\linewidth]{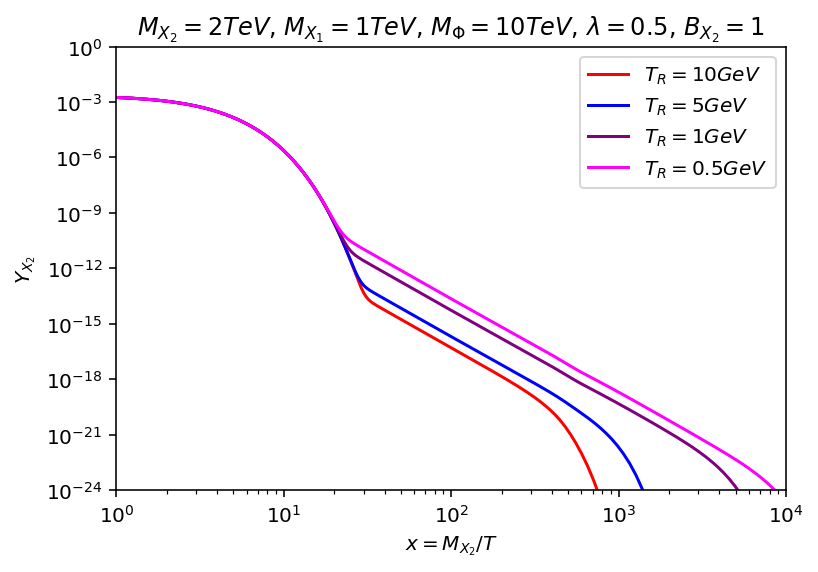}}
\subfloat{\includegraphics[width=0.5\linewidth]{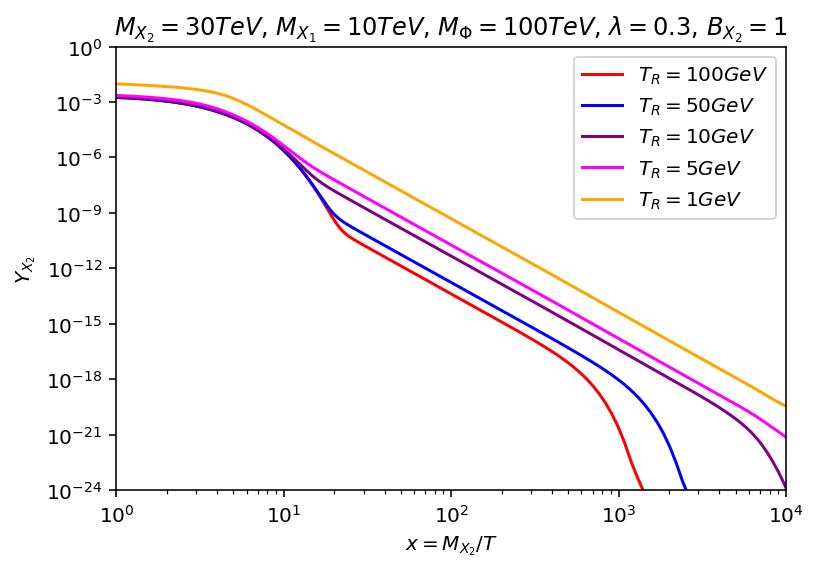}}\\
\subfloat{\includegraphics[width=0.5\textwidth]{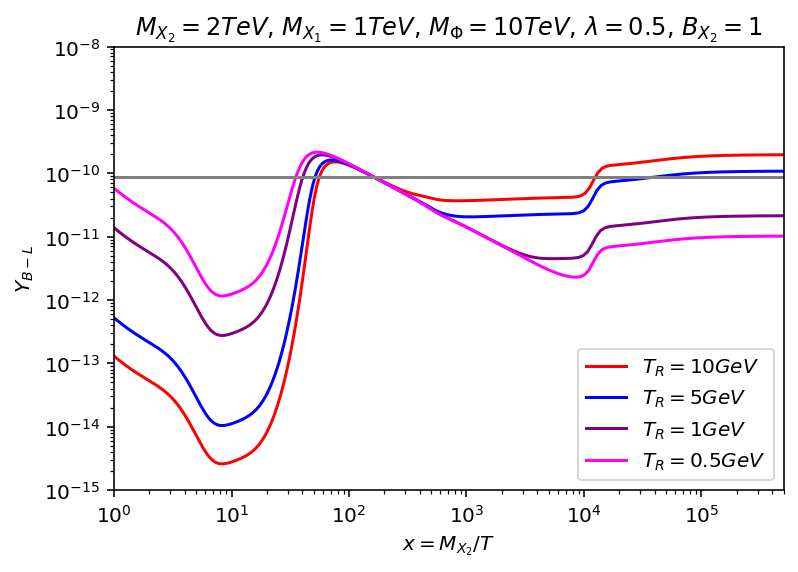}}
    \subfloat{\includegraphics[width=0.5\textwidth]{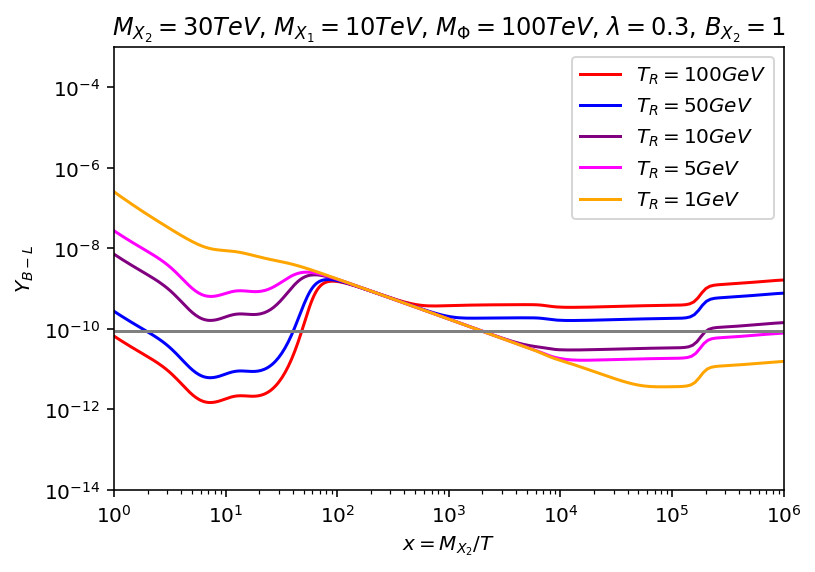}}
    \caption{\footnotesize{Evolution of $Y_{X_2}$ (upper row) and $Y_{B-L}$ (lower row) in the scenario of non-thermal production of the mother particle $X_2$ (see Sec.~\ref{sec:non-thermal_baryogen}). The two columns refer to different assignations of the $(\lambda,M_{X_1},M_{X_2},M_\Phi)$ set reported on top of each panel. In each panel the different colored curves correspond to different values of the reheating temperature, as reported in the panels themselves. In all cases the assignation $B_{X_2}=1$ has been considered. In the plots describing $Y_{B-L}$, the horizontal gray line indicates the value corresponding to the experimental determination of the baryon abundance.}}
    \label{fig:plot_solution_boltzmann}
\end{figure}

\begin{figure}
    \centering    %\includegraphics[width=0.5\linewidth]{figures_draft/pBLTb6b.pdf}
    \subfloat{\includegraphics[width=0.5\linewidth]{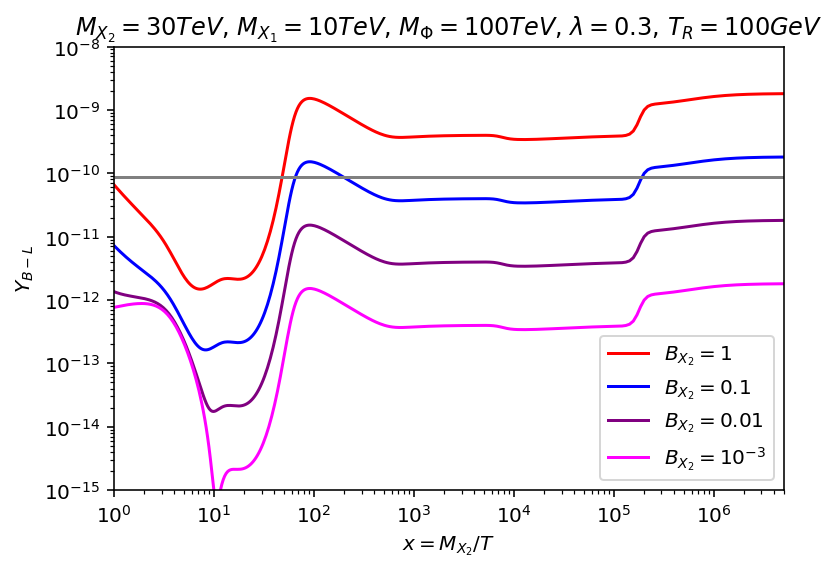}}
    \subfloat{\includegraphics[width=0.5\linewidth]{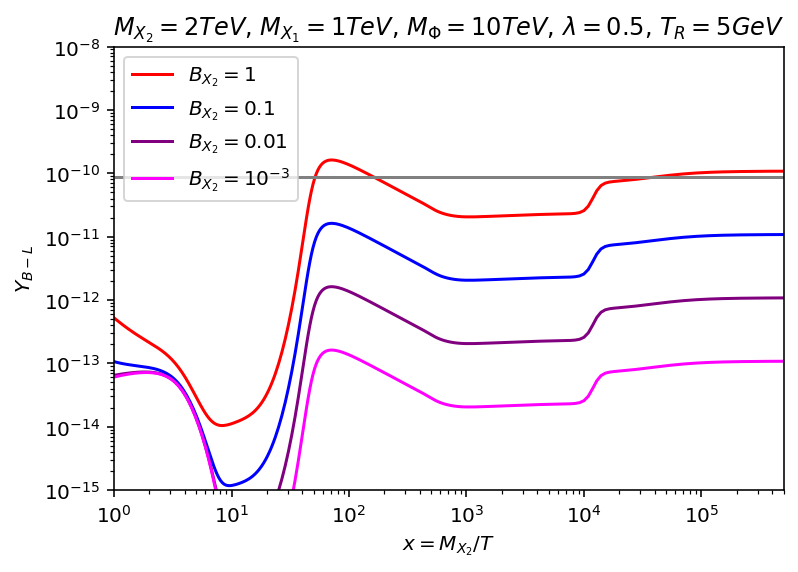}}
    \caption{\footnotesize{Evolution of $Y_{B-L}$ as function of $x=M_{X_2}/T$ for the same benchmarks already considered in Fig.~\ref{fig:plot_solution_boltzmann}. In this case, the value of $T_R$ has been kept fixed to 100~GeV (left panel) and 5~GeV (right panel). In each plot the different curves refer to different values of $B_{X_2}$, reported in the legend of each plot.}}
    \label{fig:plot_solution_boltzmann_bis}
\end{figure}

Figs.~\ref{fig:plot_solution_boltzmann} and~\ref{fig:plot_solution_boltzmann_bis}  show the numerical solutions of the system of Boltzmann's equations for some sample benchmark assignations of the model parameters. More specifically, Fig.~\ref{fig:plot_solution_boltzmann} shows the evolution of $Y_{X_2}$ (upper row) and $Y_{B-L}$ (lower row) as function of $x=M_{X_2}/T$, fixing $B_{X_2}=1$ and considering different values of $T_{R}$. The two columns of the figure refer to different assignations of the $(\lambda,M_\Phi,M_{X_1},M_{X_2})$ set. In the lower-row  panels, the value of $Y_{B-L}$ corresponding to the experimentally favored value of the baryon abundance is also shown for comparison as a gray solid line. 

For both the considered benchmarks, the final value of $Y_{B-L}$ (i.e. the plateau at $x \gg 1$) appears to increase with the value of the reheating temperature.  
This result can be understood in terms of the following analytical approximation: a fraction proportional to $B_{X_2}$ of the energy initially stored in the $\Psi$ field is directly converted into $X_2$, with negligible contribution from their annihilation processes. In the instantaneous reheating approximation,
\begin{equation}
\label{eq:X2_Yield}
    Y_{X_2}(T_{R})=\frac{n_{X_2}(T_{R})}{s(T_{R})}\simeq \frac{3}{4} B_{X_2}\frac{T_{R}}{M_\Psi}\,.
\end{equation}
The particle $X_2$ subsequently decays, generating the $B-L$ asymmetry which, in case of negligible wash-out effects, can be estimated as:
\begin{equation}
    Y_{B-L}=\epsilon_{\CP}B_{X_2,\slashed{B}}Y_{X_2}(T_{R})
\end{equation}
where
\begin{equation}
    \epsilon_{\CP}=\frac{\Delta\Gamma}{\Gamma_r}=\frac{\lambda^2}{520\pi}\frac{M_{X_2}^2}{M_\Phi^2}\,,
    % \epsilon_{\rm CP}=\frac{\Gamma_{X_r \rightarrow qqq}-\Gamma_{X_r \rightarrow \bar q \bar q \bar q}}{\Gamma_{X_r \rightarrow qqq}+\Gamma_{X_r \rightarrow \bar q \bar q \bar q}}=\frac{\lambda^2}{520\pi}\frac{M_{X_2}^2}{M_\Phi^2}
\end{equation}
while $B_{X_2,\slashed{B}}$ is the branching fraction of the baryon number-violating decays of $X_2$. From Eqs.~\eqref{eq:GammaBp}-\eqref{eq:GammaBv}, one can easily see $B_{X_2,\slashed{B}} \simeq 1$. 
Combining the previous equations, one can estimate the $B-L$ yield as: 
\begin{equation}
\label{eq:YBLapprox}
%\begin{aligned}
    Y_{B-L} \approx 1.2 \times 10^{-11} \lambda^2 B_{X2}  {\left(\frac{M_{X_2}}{100\,\mbox{GeV}}\right)}^2 {\left(\frac{1\,\mbox{TeV}}{M_\Phi}\right)}^2\left(\frac{T_{R}}{1\,\mbox{GeV}}\right)\left(\frac{10^6\,\mbox{GeV}}{M_\Psi}\right)
%    \end{aligned}
\end{equation}
Eq.~\eqref{eq:YBLapprox} shows that, for fixed $T_{R}$, the $B-L$ abundance is proportional to $B_{X_2}$. We have verified this result in Fig.~\ref{fig:plot_solution_boltzmann_bis} by plotting the $Y_{B-L}$, again as function of $x$, for a fixed value of $T_R$ and different values of $B_{X_2}$.

%Fixing the value of $B_{X_2}$ to 1, in Fig.~\ref{fig:plot_YBL_TR} we show the dependence of $Y_{B-L}$ on the reheating temperature and the coupling $\lambda$. The two panels of the figure show, as horizontal gray line, the value of $Y_{B-L}$ matching the experimentally favored value of the baryon asymmetry. The latter lines should be compared with the height of the plateu at $x \gg 1$, corresponding to the final value of $Y_{B-L}$, of the colored curves. When the latter falls above the gray line, the considered model assignation correspond to a value of $Y_{B-L}$ exceeding the experimental determination. The latter can be matched by considering a lower value of $B_{X_2}$ (see also fig. \ref{fig:plot_solution_boltzmann_bis}). On the contrary when the colored curves falls, in the region $x \gg 1$, below the gray one, the parameter assignation is not viable as $b_{X_2}=1$ is considered.  
%For some choices of these parameters, in particular the one displayed in the left panel of the figure, our model is able to produce the observed abundance of baryon asymmetry.

\begin{figure*}
    \centering
\subfloat{\includegraphics[width=0.3\textwidth]{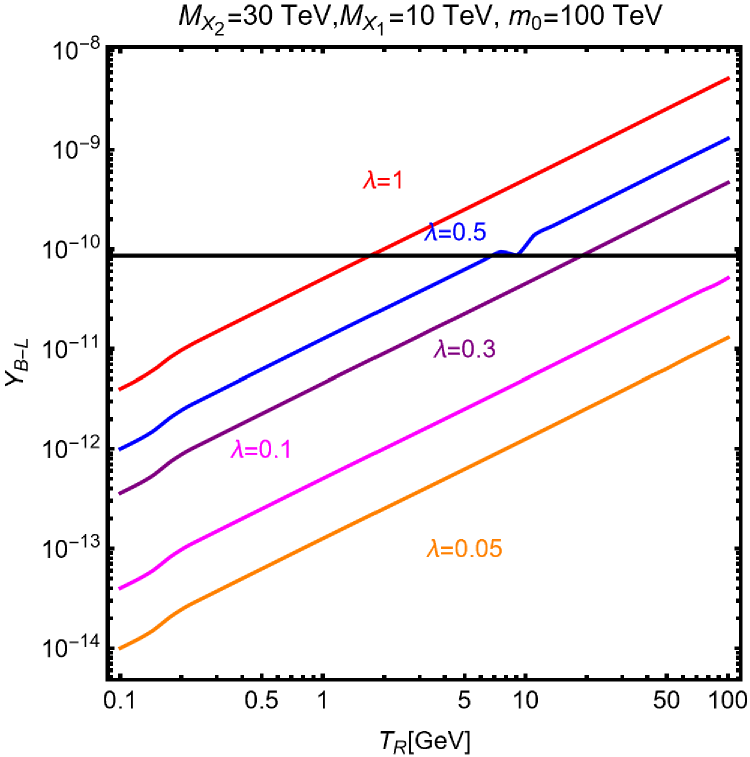}}
\subfloat{\includegraphics[width=0.3\textwidth]{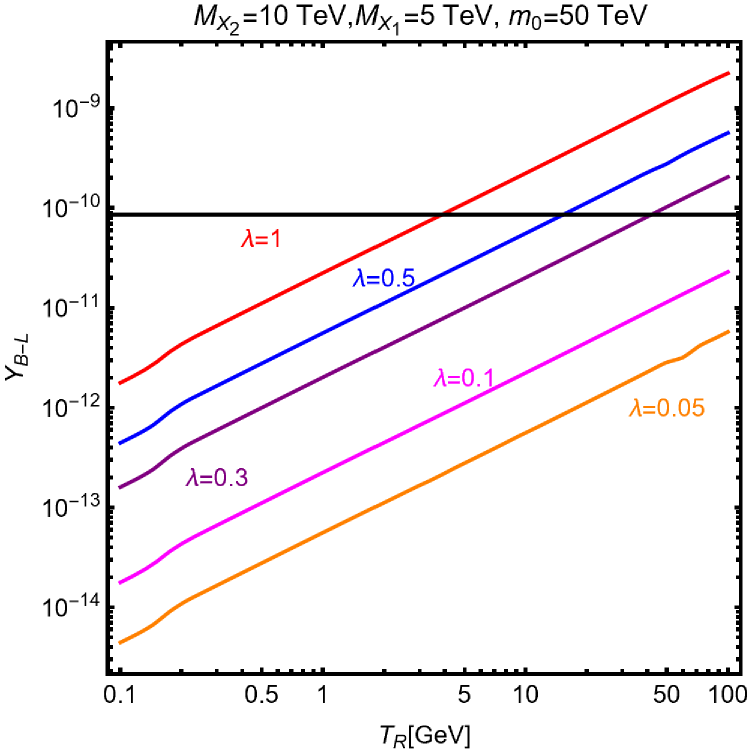}}
\subfloat{\includegraphics[width=0.3\textwidth]{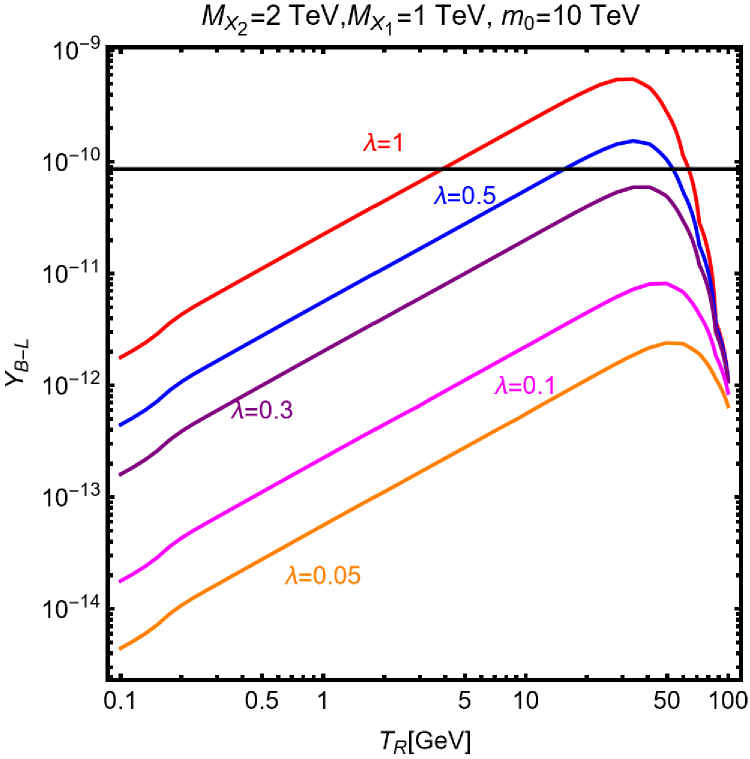}}\\
\subfloat{\includegraphics[width=0.3\textwidth]{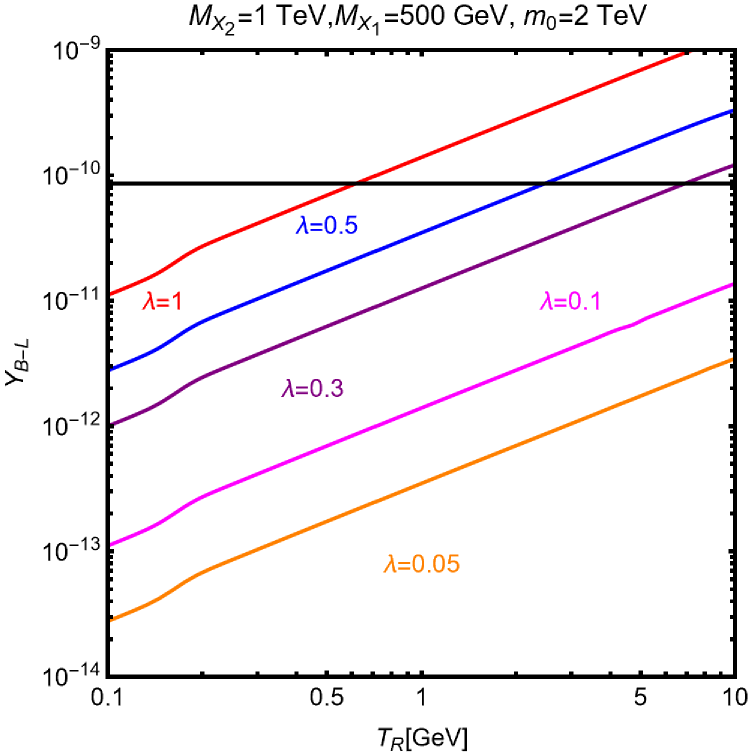}}
\subfloat{\includegraphics[width=0.3\textwidth]{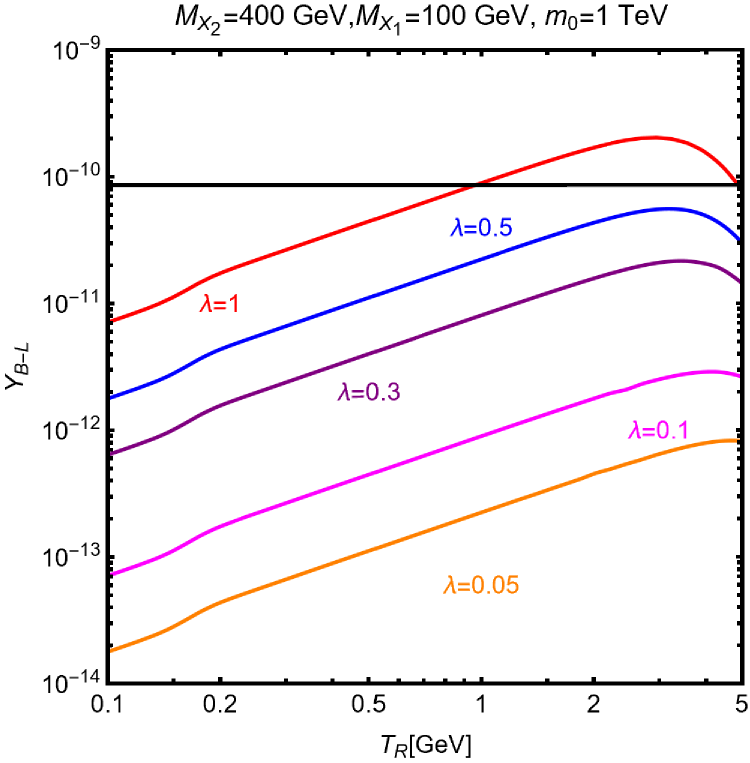}}
\subfloat{\includegraphics[width=0.3\textwidth]{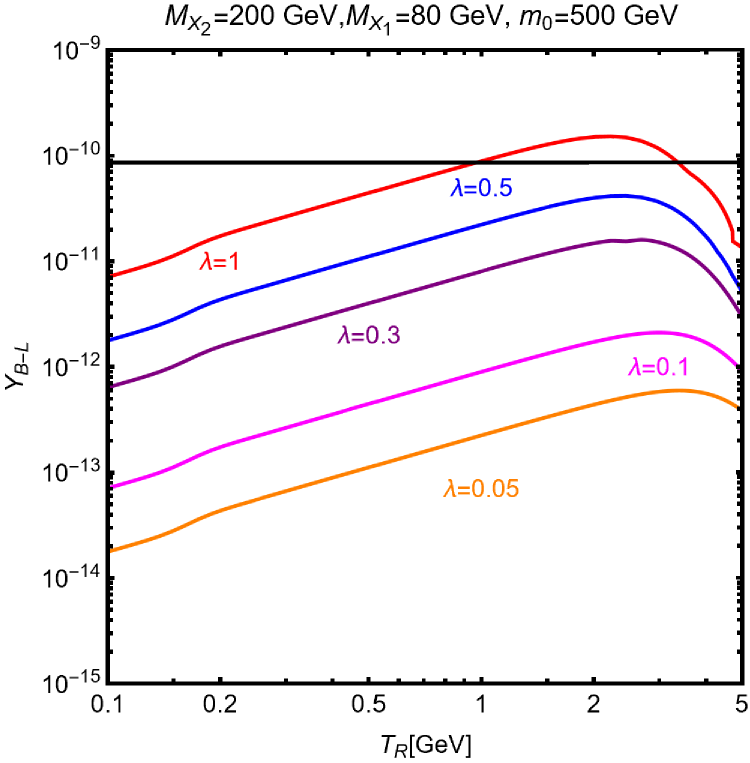}}
    \caption{\footnotesize{Final ($x\gg1)$ value of $Y_{B-L}$ as function of the reheating temperature for different assignations of the coupling $\lambda$. The different panels correspond to different assignations of the model parameters reported on top of each panel. In all cases we set $B_{X_2}=1$. In each panel the black horizontal line represent the value of $Y_{B-L}$ corresponding to the experimentally favored value of the baryon abundance.}}
    \label{fig:plot_YBL_TR}
\end{figure*}

Fig.~\ref{fig:plot_YBL_TR} displays a more systematic analysis focused on the behavior of $Y_{B-L}$ with the reheating temperature for different assignations of the coupling $\lambda$. The different panels show a series of benchmarks assignations of the $(M_\Phi,M_{X_1},M_{X_2})$ parameters which cover a mass range for the $X_{1},\,X_2$ particles from $\mathcal{O}(100\,\mbox{GeV})$ to $\mathcal{O}(10\,\mbox{TeV})$. Each panel shows a series of colored curves, each one corresponding to a value of $\lambda$ between 0.05 and 1. In each case, the parameter $B_{X_2}$ was set equal to 1.
We notice that $Y_{B-L}$ increases with the reheating temperature $T_R$, in agreement with the analytical expression~\eqref{eq:YBLapprox}, until it reaches a turnover point for high reheating temperature
% By looking at the individual curves, we notice that in the majority of the plots, $Y_{B-L}$ increases with the reheating temperature $T_R$ in agreement with the analytical expression~\eqref{eq:YBLapprox}. Some plots show nevertheless deviations from this trend 
\footnote{In the benchmarks with $M_{X_2}=1,10,30\,\mbox{TeV}$ the decrease of $Y_{B-L}$ does not appear because of chosen range of the plot.}. This occurs when the latter % reheating temperature
approaches the temperature of the standard thermal freeze-out of the mother particle $X_2$, which is approximately $M_{X_2}/20$. In such a case, the non-thermal production becomes less effective as the decay products of $\Psi$ thermalize with the primordial plasma. When the reheating temperature exceeds the standard thermal freeze-out temperature of $X_2$, the scenario is undistinguishable from the conventional thermal production for $X_2$.

The horizontal black lines in Fig.~\ref{fig:plot_YBL_TR} represent the experimentally favored value of $Y_{B-L}$. A considered assignation of $(\lambda,M_{X_1},M_{X_2},M_\Phi)$ hence corresponds to a viable baryogenesis for the values of the reheating temperature at which the colored lines cross the black line. In the regions in which a curves lie above the horizontal line, the baryon abundance is overproduced. Nevertheless, viable baryogenesis is achievable by lowering the value of $B_{X_2}$. On the contrary, if a curve lies always below the horizontal line, the corresponding parameter assignation cannot account for the experimental value of the baryon abundance, so one would need to extend the model to accommodate additional sources of baryon asymmetry.

\section{Including DM}
\label{sec:including_DM}
\begin{figure}
    \centering
\subfloat{\includegraphics[width=0.5\linewidth]{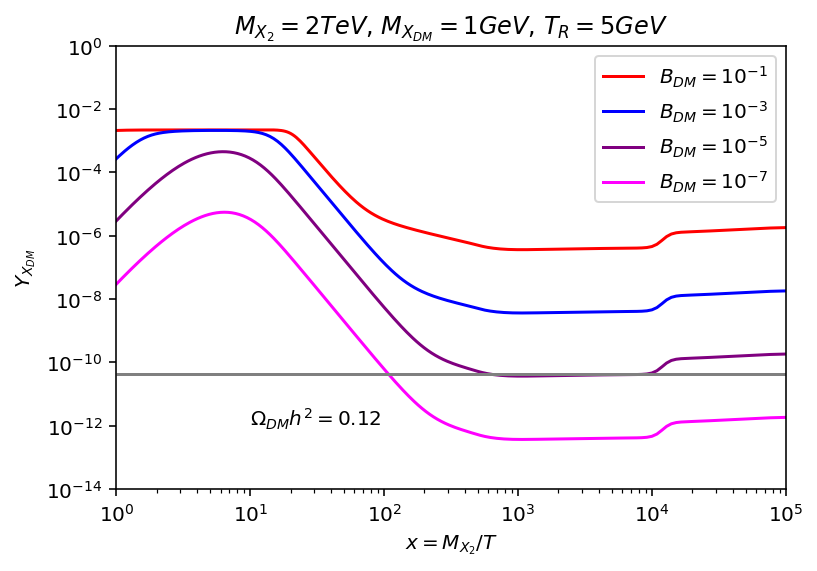}}
\subfloat{\includegraphics[width=0.5\linewidth]{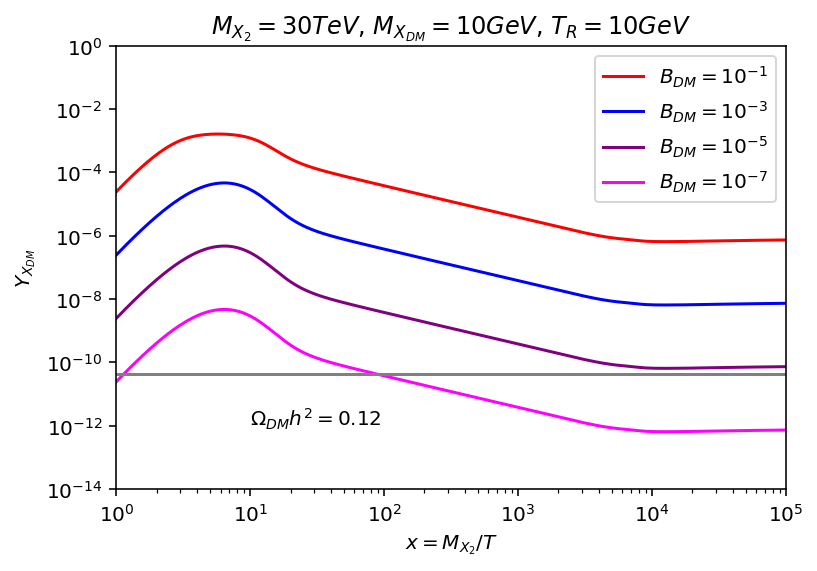}}
    \caption{\footnotesize Evolution of the DM abundance as function of $x=M_{X_2}/T$ in the scenario in which the DM is solely produced by the decay of the $X_2$ state (Eq.~\eqref{eq: boltz_NDM_X2decay}. The different colored curves correspond to different values of $B_{\rm DM}$. The gray horizontal line correspond to the experimentally favored value of the DM relic density. All the curves refer to the same assignation of the $(M_{X_2},T_{R},M_{X_{\rm DM}})$ set, reported on top of the panel. The other parameters have been set as $M_\Phi=100\,\mbox{TeV}$, $M_{X_1}=10\,\mbox{TeV}$, $B_{X_2}=1$ and $\lambda=0.3$.}
    \label{fig:plot_dm_1}
\end{figure}

The most economical way to incorporate a DM candidate in our setup consists in extending the particle spectrum with an additional state~\footnote{The state $X_1$ needs to be unstable, otherwise the coupling needed for the generation of $\epsilon_{\CP}$ would be forbidden.}, here dubbed $X_{\rm DM}$. This option can be accommodated for in a Supersymmetric realization considering gravitino DM~\cite{Arcadi:2013jza,Arcadi:2015ffa} and creates a rather elegant connection between the production of DM and the baryon asymmetry, occurring from the decay of the same mother particle. 
Here for simplicity we consider instead the DM candidate to be a Majorana fermion coupled with the particle $X_2$ responsible for the production of the baryon asymmetry. Sticking to a simplified setup and performing the usual change of variable $N_{X_{\rm DM}} = n_{X_{\rm DM}} a^3$, the evolution of the DM abundance can be described by a simple Boltzmann's equation of the form:
\begin{equation}
\begin{aligned}
\label{eq: boltz_NDM_X2decay}
    & \frac{d N_{{X_{\rm DM}}}}{dA}=\frac{A^{1/2}\Lambda^{-3/2}}{\mathcal{H}}\langle \Gamma_{\rm DM} \rangle  \left(N_{X_2}-\frac{N_{X_{\rm DM}}}{N_{X_{\rm DM},\eq}}N_{X_2,\eq}\right)\,,
    \end{aligned}
\end{equation}
coupled to Eqs.~\eqref{eq:BoltzNT_Psi}-\eqref{eq:BoltzNT_BL}. $\Gamma_{\rm DM}$ represents the decay rate of $X_2$ into the DM particle. Notice that in this setup, the baryon abundance would be possibly reduced by a factor $(1-B_{\rm DM})$, $B_{\rm DM}$ being the decay branching fraction of $X_2$ into DM. To limit this possibility we take $B_{\rm DM}\leq 0.1$ so that, in good approximation, we can set
\begin{equation}
    \Gamma_{\rm DM} \simeq B_{\rm DM} \left(\Gamma_{21}+\Gamma_2\right) 
\end{equation}
where $\Gamma_{21}$ and $\Gamma_{2}$ are the rates defined in Eqs.~\eqref{eq:GammaBp} and \eqref{eq:GammaBv}. The DM abundance is then simply related to the one of $X_2$ by $Y_{X_{\rm DM}}=B_{X_{\rm DM}} Y_{X_2}$. Using then Eq.~\eqref{eq:X2_Yield} and $\Omega_{\rm DM}h^2 \simeq 2.82 \times 10^8 M_{X_{\rm DM}} Y_{X_{\rm DM}}$ \cite{Gondolo:1990dk},
\begin{equation}
\label{eq:DM_1}
%\begin{aligned}
    \Omega_{X_{\rm DM}}h^2 \approx 0.47 \times 10^4 B_{X_{\rm DM}}B_{X_2}\left(\frac{M_{X_{\rm DM}}}{10\,\mbox{GeV}}\right) \left(\frac{T_{\rm R}}{1\,\mbox{GeV}}\right)\left(\frac{10^6\, \mbox{GeV}}{M_\Psi}\right)\,.
 %   \end{aligned}
\end{equation}
A sample solution of the Boltzmann's equation for the DM is illustrated in Fig.~\ref{fig:plot_dm_1}.
In agreement with the analytical expression \eqref{eq:DM_1}, DM is produced very efficiently so that the correct value of the relic density is obtained for a value of $B_{\rm DM}$ between $10^{-6}$ and $10^{-5}$.

An alternative implementation of DM can be performed, along the same lines as~\cite{Pierce:2019ozl}, by considering an additional particle sector characterized by a new abelian gauge symmetry dubbed $U(1)_D$. The DM state $X_{\rm DM}$ is now a vector-like fermion charged under the new symmetry interacting with the corresponding gauge boson, dubbed $Z'$ as customary. The DM can be related to the sector responsible for the baryogenesis by introducing a scalar field $\Sigma$ also charged under $U(1)_D$ symmetry so that a Yukawa-like coupling between the $X_2$, $X_{\rm DM}$ and $\Sigma$ states can be written. In summary, the Lagrangian responsible for DM interactions can be written as:
\begin{equation}
    \mathcal{L}_{\rm DM}=g_D \bar X_{\rm DM}\gamma^\mu X_{\rm DM}Z'_{\mu}+y_D \bar X_{\rm DM} X_2 \Sigma +\mbox{h.c.}\,
\end{equation}
where $g_D$ and $y_D$ are coupling strengths.
The $U(1)_D$ symmetry ensures DM stability as long as $M_{X_2}>M_{\rm DM}$ and $M_{\Sigma}>M_{\rm DM}$. We further assume $M_{\Sigma}=M_\Phi \gg M_{\rm DM}$ so that the decay of $X_2$ into DM is kinematically forbidden, and $y_D=g_D$ for simplicity. The Yukawa coupling is nevertheless responsible of the $X_2 X_2 \rightarrow X_{\rm DM} X_{\rm DM}$ annihilation process (the corresponding cross-section is given by an analogous expression as the one in the first row of Eq.~\eqref{eq:ann_processes}), which hence relates the DM and $X_2$ abundances. The DM abundance is also determined by annihilation processes into $Z'Z'$, if $M_{\rm DM} > M_{Z'}$, and possibly into SM pairs, if kinetic mixing between the new gauge boson and the hypercharge one is accounted for. We also assume the DM to be non-thermally produced by the decay of $\Psi$.
The Boltzmann's equation tracking the DM abundance can be hence written as:
    \begin{align}
   \begin{split}
    \label{eq:BoltzNT_DM}
         \frac{dN_{X_{\rm DM}}}{dA}&=-\frac{A^{-5/2}\Lambda^{3/2}}{\mathcal{H}} \bigg[\langle \sigma_{\rm DM} v \rangle \left(N_{X_{\rm DM}}^2-N_{X_{\rm DM},\eq}^2\right)- \langle \sigma_{2X} v \rangle \left(N_{X_2}^2-\frac{N_{X_{\rm DM}}^2}{N_{X_{\rm DM},\eq}^2}N_{X_2,\eq}^2\right)\bigg]\\
        & + \Lambda\frac{B_{\rm DM}}{M_\Psi}\Gamma_\Psi \xi_\Psi \frac{A^{1/2}\Lambda^{-3/2}}{\mathcal{H}}\,,
   \end{split}
    \end{align}
where $\langle \sigma_{2X}v \rangle \equiv \langle \sigma_{X_2 X_2 \rightarrow X_{\rm DM}X_{\rm DM}} v \rangle$ while 
%$\langle \sigma_{\rm DM} v \rangle \equiv \langle \sigma v \rangle \left(X_{\rm DM} X_{\rm DM} \rightarrow Z^{'}Z^{'}\right)+\langle \sigma v \rangle \left(X_{\rm DM}X_{\rm DM} \rightarrow SM SM\right)$ 
$\langle \sigma_{\rm DM} v \rangle \equiv \langle \sigma_{X_{\rm DM} X_{\rm DM} \rightarrow Z^{'}Z^{'}} v \rangle $ (we neglect for simplicity the presence of kinetic mixing; analytical expressions for the cross-section can be found for example in \cite{Arcadi:2024ukq}).

\begin{figure}
    \centering    \subfloat{\includegraphics[width=0.5\linewidth]{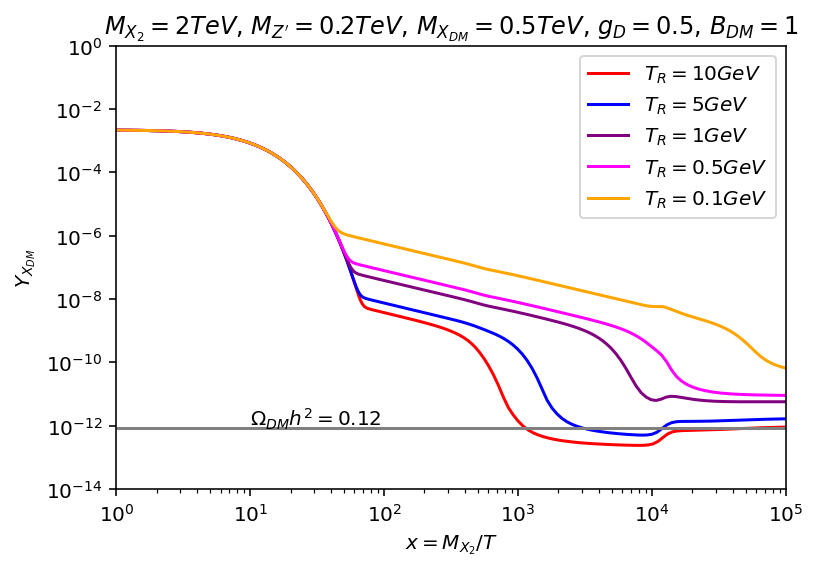}}
\subfloat{\includegraphics[width=0.5\linewidth]{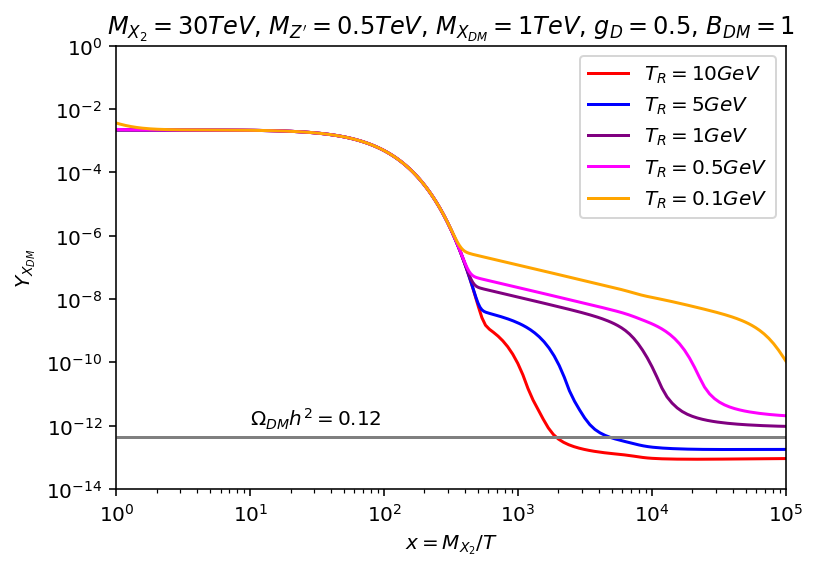}}
    \caption{\footnotesize{Evolution of the DM abundance as function of $x=M_{X_2}/T$ in the dark matter implementation with an extra $U(1)$ (Eq.~\eqref{eq:BoltzNT_DM}). The different curves correspond to different values of the reheating temperature for the fixed values $M_{X_{\rm DM}}=500\,\mbox{GeV}$, $M_{Z'}=200\,\mbox{GeV}$, $g_D=0.5$, $M_{X_1}=1\,\mbox{TeV}$, $M_{X_2}=2\,\mbox{TeV}$, $M_\Phi=10\,\mbox{TeV}$. In all cases we set $B_{\rm DM}=1$.}}
    \label{fig:plot_dm_Zp1}
\end{figure}

Eq.~\eqref{eq:BoltzNT_DM} should be coupled with the system~\eqref{eq:BoltzNT_Psi}-\eqref{eq:BoltzNT_BL} (the term proportional to $\langle \sigma_{2X}v\rangle$ should be included as well in the equation for $X_2$). We show in Fig.~\ref{fig:plot_dm_Zp1} the numerical solution of this system for two assignations of the set $(\lambda,M_{X_1},M_{X_2},M_\Phi,M_{X_{\rm DM}},M_{Z'})$, 
%namely $(0.5, 1\,\mbox{TeV}, 2\,\mbox{TeV},10\,\mbox{TeV}, 500\,\mbox{GeV},200\,\mbox{GeV})$  and $(0.3, 10\,\mbox{TeV}, 30\,\mbox{TeV},100\,\mbox{TeV}, 1\,\mbox{TeV},500\,\mbox{GeV})$ \AM{Do we have to write these numbers explicitly here? They are badly rendered in the pdf}, 
and for different values of the reheating temperature, corresponding to the different colored curves in each plot. In all cases, we considered $B_{X_2}=B_{\rm DM}=1$. Fig.~\ref{fig:plot_dm_Zp1} also evidences, through horizontal gray lines, the values of $Y_{X_{\rm DM}}$ corresponding to the experimental value of the DM relic density. % As evident, the contours of the comoving DM abundance are rather different from the ones of $Y_{B-L}$, for the same assignation of $(\lambda,M_{X_1},M_{X_2})$.
We notice that, at early times the comoving abundance of DM matches the value expected for an equilibrium distribution function for the DM. At intermediate times, $Y_{X_{\rm DM}}$ decreases smoothly with $x$ and then features a second sharp drop before getting to the final asymptotic value at late times. Most importantly, the DM relic density decreases as the value of the reheating temperature decreases. This behavior of the solution of the Boltzmann's equation is compatible with the analytical solution dubbed in~\cite{Gelmini:2006pq} ``non-thermal production with chemical equilibrium'', which can be expressed schematically as:
\begin{equation}
    \Omega_{X_{\rm DM}}h^2 \approx \frac{T_{\rm s.f.o}^3}{T_{R}^3}\Omega_{X_{\rm DM}}^Th^2
\end{equation}
with $\Omega_{X_{\rm DM}}^T h^2$ representing the relic density which would be obtained for the same DM masses and coupling, but assuming only thermal production in a standard cosmological history. $T_{\rm s.f.o.}$ represents the freeze-out temperature in the standard thermal paradigm.
The behavior, just described, of the solution of the Boltzmann's equation for DM is due to the presence of the very efficient annihilation process into $Z'Z'$.

\begin{figure}
    \centering
\subfloat{\includegraphics[width=0.5\linewidth]{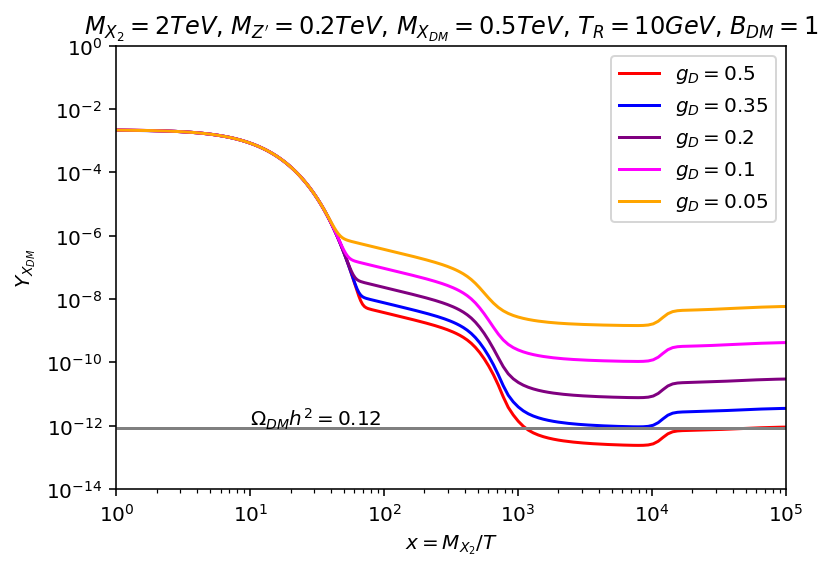}}
\subfloat{\includegraphics[width=0.5\linewidth]{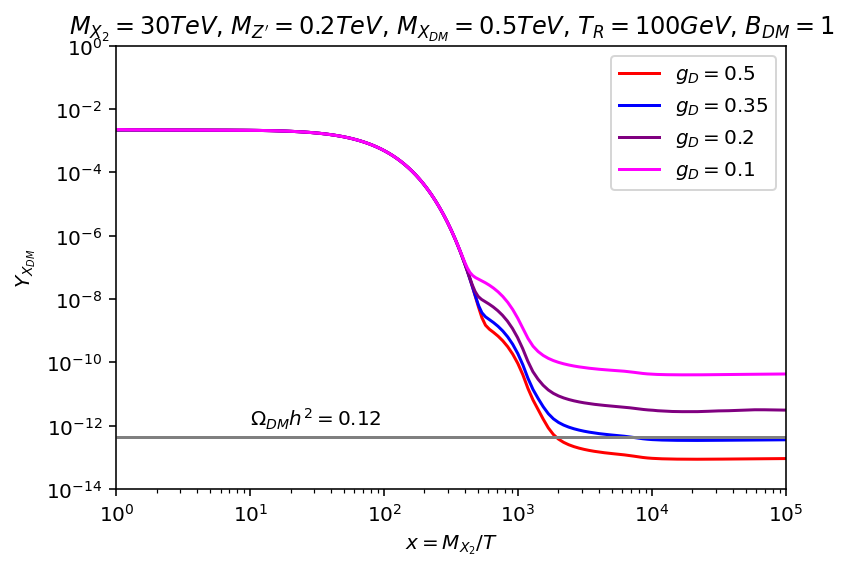}}
    \caption{\footnotesize{Evolution of the DM abundance, in the implementation with extra $U(1)_D$, as function of $x=M_{X_2}/T$ for two benchmark assignations of the model parameters. In each plot a specific value of the reheating is considered, namely $T_R=10\,\mbox{GeV}$ (left panel) and $T_R=100\,\mbox{GeV}$ (right panel), and the values $B_{X_2}=1$ and $B_{\rm DM}=1$ have been considered. The different colored curves correspond to different values of $g_D$. In both panels the values of the DM and $Z'$ masses have been set to, respectively, $500\,\mbox{GeV}$ and $200\,\mbox{GeV}$.}}
    \label{fig:plot_dm_Zp2}
\end{figure}

The dependence of the DM relic density on the model parameters is further investigated in Fig.~\ref{fig:plot_dm_Zp2}. The two panels show two assignations of the $(\lambda,M_{X_1},M_{X_2},M_\Phi, T_{\rm R},B_{X_2})$ set complying with viable baryogenesis (see previous section). These assignations are combined with the choices $(500\,\mbox{GeV},200\,\mbox{GeV})$ and $(1\,\mbox{TeV},0.5\,\mbox{GeV})$ of the $(M_{X_{\rm DM}},M_{Z'})$ pair. The panels of Fig.~\ref{fig:plot_dm_Zp2} show the DM comoving abundance $Y_{X_{\rm DM}}$, for the parameter assignations just illustrated, and different values of $g_D$. In all cases we considered $B_{\rm DM}=1$. As customary, the gray horizontal line in each plot allows to point out the parameter assignation accounting for the correct DM relic density. In agreement with the analytical estimate, the DM abundance decreases with the increase of the DM gauge coupling. The experimentally-favored value is reached, for the considered benchmarks, for the values $g_D=0.5$ (left panel of Fig.~\ref{fig:plot_dm_Zp2}) and $g_D=0.35$ (right panel of Fig.~\ref{fig:plot_dm_Zp2}). As already pointed out, the chosen benchmarks also allow to accommodate for the correct value of the baryon abundance.

\section{Conclusions}
\label{sec:conclusions}
We have proposed a framework for the combined solution of the baryogenesis and DM puzzles within a non-standard cosmological history. We assumed the presence of an exotic state $\Psi$, responsible for an early matter-dominated phase in the history of the Universe and for the non-thermal production of a WIMP-like particle, which in turn generates the matter-antimatter asymmetry via $B$- and $\CP$-violating annihilation and decay processes. Leaving for a future work an extensive numerical study, we provide here a successful sample case-study represented by a simplified model containing the minimal ingredients to realize the baryogenesis mechanism under scrutiny. These are the presence, for the mother particle $X_2$, of both $B-$violating interactions, mediated by a BSM scalar field $\Phi$, and $B-$conserving interactions, requiring the presence of an additional particle $X_1$ lighter than $X_2$. We have numerically solved the Boltzmann's equations for assignations of the masses of $X_{1}$ and $X_{2}$ ranging from $\mathcal{O}(100\,\mbox{GeV})$ to $\mathcal{O}(10\,\mbox{TeV})$, complying with a viable baryogenesis. We have then proposed two scenarios to incorporate DM in this setup. The first one simply includes an additional particle produced via the decays of $X_2$. The second, richer scenario assumes the presence of a dark sector featuring an extra abelian symmetry. The DM candidate is non-thermally produced by the decays of the $\Psi$ state. A connection between the dark sector and the one accounting for baryogenesis is nevertheless present, ensured by the $X_2 X_2 \leftrightarrow X_{\rm DM} X_{\rm DM}$ processes.

%%%%%%%%%%%%%%%%%%%%%%%%%%%%%%%%%%%%%%%%%

\bibliography{biblio}
\end{document}